\documentclass[floats,aps,showpacs,amssymb,prd,onecolumn,secnumarabic%
,tightenlines%
,nofootinbib]{revtex4}
\usepackage{amssymb}
\usepackage{graphics}
\usepackage{graphicx}
\usepackage{amsmath}
\usepackage{amsfonts}
\usepackage{bm}

\begin{document}

\title{PeV IceCube signals and Dark Matter relic abundance in modified cosmologies}%

\author{G. Lambiase$^{a,b}$, S. Mohanty$^c$ and An. Stabile$^{a,b}$}

\affiliation{$^a$Dipartimento di Fisica ''E.R. Caianiello'' Universit\`a di Salerno, I-84084 Fisciano (Sa), Italy,}
\affiliation{$^b$INFN - Gruppo Collegato di Salerno, Italy.}
\affiliation{$^c$Physical Research Laboratory, Ahmedabad, 380009, India}

\date{\today}
\def\be{\begin{equation}}
\def\ee{\end{equation}}
\def\al{\alpha}
\def\bea{\begin{eqnarray}}
\def\eea{\end{eqnarray}}

\renewcommand{\theequation}{\thesection.\arabic{equation}}
\begin{abstract}

The discovery by the IceCube experiment of a high-energy astrophysical neutrino flux with energies of the order of PeV, has opened new scenarios in astroparticles physics. A possibility to explain this phenomenon is to consider the minimal models of Dark Matter (DM) decay, the 4-dimensional operator $\sim y_{\alpha \chi}\overline{{L_{L_{\alpha}}}}\, H\, \chi$, which is also able to generate the correct abundance of DM in the Universe.
 Assuming that the cosmological background evolves according to the standard cosmological model, it follows that the rate of DM decay $\Gamma_\chi \sim |y_{\alpha \chi}|^2$ needed to get the correct DM relic abundance ($\Gamma_\chi\sim 10^{-58}$) differs by many orders of magnitude with respect that one needed to explain the IceCube data ($\Gamma_\chi\sim 10^{-25}$), making the four-dimensional operator unsuitable. In this paper we show that assuming that the early Universe evolution is governed by a modified cosmology, the discrepancy between the two the DM decay rates can be reconciled, and both the IceCube neutrino rate and relic density can be explained in a minimal model.
\end{abstract}


\maketitle

\section{Introduction}
\setcounter{equation}{0}

The IceCube Collaboration,  in its 4-year dataset \cite{[2],[3]},  had reported three neutrino-induced cascade events with energies $\sim 1 $~PeV \cite{[3]}.
First candidates for the generation of such neutrino high energy events were various astrophysical sources  \cite{[14],[15],[16],Sahu:2014fua}.
However after the analysis of  seven years of data of muon tracks, the  IceCube reports \cite{Aartsen:2016oji} that there is  no clear correlations with the known astrophysical hot-spots like the known supernova remnants (SNR) or Active Galactic Nuclei (AGN). In the light of no identification of the astrophysical  sources of IceCube neutrinos it has been proposed that the neutrinos could arise from the decay of PeV mass Dark Matter (DM)  \cite{[17],[21],[22],[23],[24],[25],[26],[27],[28],[29],[30],[31],[32],[33],[34],[35]} (the DM if boosted could may be of lower mass  also \cite{[36], [37]}).

Unitarity bounds on the cross section of the DM \cite{Griest:1989wd, Beacom:2006tt} ruled out the possibility of a thermal relic density of PeV scale DM whose annihilation or decay could produce the IceCube neutrinos. Other production mechanisms which have been invoked for the PeV mass dark matter relic density are a secluded sector [39], freeze-out with resonantly enhanced annihilations \cite{[40]}, or freeze-in \cite{[25],[33],[41],[52]}. To explain IceCube events with DM of mass of $\sim$ PeV with lifetime of $\tau \sim 10^{28}$ sec is required \cite{[17],[21]}. If the neutrino flux at IceCube were to be from DM annihilation into neutrinos then the same then the same decay rate $n_\chi/\tau_\chi$ must be equal to the annihilation rate $n_\chi^2 \langle \sigma v\rangle $ which implies that $\langle \sigma v \rangle \simeq (n_\chi \tau_\chi)^{-1} \simeq 10^{-17} cm^3/sec $ which is again ruled out from unitarity constrains \cite{Griest:1989wd, Beacom:2006tt}. This implies that PeV DM decay  with lifetime $10^{28}$ sec is the prefered mechanism for explaining the IceCube neutrinos,  if at all they originate from DM.

Chianese and Merle \cite{merle} raised the question of whether it is possible to explain both the PeV DM relic density and the decay rate required for IceCube with one operator. The minimal DM-neutrino dimension four interaction $ y\, \bar L\cdot H \chi$ of DM decay is able to produce the correct DM abundance by freeze-in for $y\sim 10^{-12}$ \cite{merle} but to get a decay lifetime of $10^{28}$ sec the value of $y \sim 10^{-29}$  \cite{[34], [35], merle}.  This implies that the minimal dimension four operator fails to account for both the PeV dark matter relic abundance and the decay rate required to explain IceCube.

The DM models analyzed in \cite{merle}, however, are based on General Relativity and the standard cosmological inflation followed by a radiation dominated era. The signal of inflation is in the cosmic microwave anisotropy spectrum while the only experimental evidence of the radiation era is the successful predictions of big-bang nucleosynthesis (BBN) which occurs around T$\sim $ 1 MeV at $t \sim 1$ sec. Observations from Type Ia Supernovae \cite{Riess}, CMB radiation \cite{Spergel}, and the large scale structure \cite{Tegmark,Eisenstein}, suggest that there are strong evidences that the present cosmic expansion of the Universe is accelerating. The latter is ascribed to the existence of Dark Energy (DE), an exotic form of energy characterized by a negative pressure  that at late times dominates over the cold and dark matter, driving the Universe to the observed accelerating phase, and new ingredients, such as DM and Dark Energy (DE), are required \cite{review1}. Both inflation and dark energy have motivated the extension of Einstein's theory to a general $f(R)$ theory \cite{DeFelice:2010aj}. For example the most successful model of inflation which is consistent with the low tensor to scalar ratio $r$ favored by experiment is the Starobinsky model  \cite{starobinsky1980} $ {\cal L}= M_P^2 R + (1/M^2) R^2 $.  The $R+R^2$ Starobinsky model can be generalized to a
$R+R^n$ model \cite{Kehagias:2013mya, girish} which  predicts a larger tensor to scalar ration than that allowed by the Starobinsky model and  which may be accessible to experimental efforts to measure the primordial B-mode polarization in the CMB. The Starobinsky model and its generalization can also be derived from Supergravity \cite{Cecotti:1987sa, Ellis:2013xoa}.
In addition to inflation  another motivation for $f(R)$ gravity models is to explain dark energy model \cite{Hu-Sawicki, starobinsky}. In addition to these cosmologically motivated generalizations of Einstein's gravity there is been many classical attempts to generalize Einstein's gravity by adding a scalar component  to the tensor metric theory  as in the Brans-Dicke model \cite{reviewBD,BransDicke}.

From a cosmological point of view, one of the consequences of dealing with the cosmology based on modified gravity is that the thermal history of particles gets  modified. This means that if the cosmological background is described by modified cosmologies, the expansion rates $H$ of the Universe can be written in terms of the expansion rate $H_{GR}$ of GR, i.e. $H(T)=A(T)H_{GR}(T)$. Here the factor $A(T)$ encodes the information about the underlying model of gravity that extend/modify GR. Typically, the factor $A(T)$ is defined in a way that the successful predictions of the BBN are preserved, so that $A(T) \neq 1$ at early time, i.e. at the pre-BBN epoch, an epoch of the Universe not directly constrained by cosmological observations, while $A(T)\to 1$ when (or before) BBN starts.

In this paper we consider generalizations  of the standard cosmology with the aim to get a consistent minimal model of PeV DM and IceCube neutrinos. In particular we calculate the freeze-in abundance of DM in a modified cosmology and show that the couplings required for obtaining the required relic abundance depend upon the modified gravity parameters. By choosing the cosmological parameters appropriates such that there is no deviation from the predictions of BBN, we can get the minimal model of decaying DM ${\cal L}=  y_\alpha\, \bar L_\alpha \cdot H \chi$($\alpha $ indicates the  mass eigenstates of the three active neutrinos, $\chi$ the DM particle, $H$ the Higgs doublet, $L_{L_{\alpha}}$ the left-handed lepton doublet, and $y_{\alpha \chi}$ the Yukawa couplings)  satisfy relic abundance and the IceCube requirements with single value of a combination of the couplings $\sum_\alpha |y_\alpha^2|$.

The paper is organized as follows. In Section 2 we recall the main topic of DM relic abundance and the IceCube data, pointing out that they cannot be consistently explained by using the 4-dimensional operator. In Section 3 we show that the latter allows to explain both the DM relic abundance and the IceCube experiment if it is assumed that after Inflation, the Universe evolution is described by modified cosmologies (instead of the standard cosmological model) at least till the BBN starts. As example of modified cosmologies, we shall consider scalar tensor theories, Brans-Dicke theory and $f({\cal T})$ theories, where ${\cal T}$ is the scalar torsion. Conclusions are given in the last Section 4.

\section{PeV neutrinos and Ice Cube data}

In this Section, we recall the main features related to DM relic abundance and IceCube data \cite{merle}.
The simplest 4-dimensional operator able to explain the IceCube high energy signal, is given by the Lagrangian density
 \begin{equation}\label{4dop}
 {\cal L}_{d=4}=y_{\alpha \chi}\overline{{L_{L_{\alpha}}}}\, H\, \chi\,, \qquad \alpha = e, \mu, \tau\,,
 \end{equation}
where $\chi$ is the DM particle that transforms as $\chi\sim (1,1,0)$ of SM, $H\sim (1, 2, +1/2)$ is the Higgs doublet, $L_{L_{\alpha}}\sim (1, 2, -1/2)$ is the left-handed lepton doublet corresponding to the generation $\alpha (=e, \mu\, \tau)$, and finally $y_{\alpha \chi}$ are the Yukawa couplings.

Following \cite{hall,merle}, we confine ourselves to  freeze-in production, i.e. the DM particles are never in thermal equilibrium since they interact very weakly, but are gradually produced from the hot thermal bath. This occurs owing to a feeble coupling to particles of the SM (at $T\gg m_\chi$), allowing to DM particles to remain in the Universe because of the smallness of the back-reaction rates and the slowness of the decay to occur. Therefore a sizable DM abundance is allowed in this model, at least until the temperature falls down to $T\sim m_\chi$ (temperatures below $m_\chi$ are such that DM particles phase-space is kinematically difficult to access).

The evolution of the DM particle is governed by the Boltzmann equation. Denoting with $Y_\chi=n_\chi/s$ the DM abundance, where $n_\chi$ is the number density of the DM particles and $s=\frac{2\pi^2}{45}g_*(T)T^3$ the entropy density ($g_*$ denotes the degrees of freedom), from the Boltzmann equation one gets
 \begin{equation}\label{Boltz}
   \frac{dY_\chi}{dT}=-\frac{1}{HTs}\left[\frac{g_\chi}{(2\pi)^3}\int C\frac{d^3p_\chi}{E_\chi}\right]\,,
 \end{equation}
where $H$ is the expansion rate of the Universe and $C$ the general collision term. For cosmological models in which is assumed that the relativistic degree of freedom are constant, i.e. $dg_*/dT=0$, the DM relic abundance assumes the form
 \begin{equation}\label{DM1}
   \Omega_{DM}h^2=\frac{2 m_\chi^2 s_0 h^2}{\rho_{cr}}
   \int_0^\infty\frac{dx}{x^2}\left(-\frac{dY_\chi}{dT}\Big|_{T=\frac{m_\chi}{x}}\right)\,,
 \end{equation}
where $x=m_\chi/T$, $s_0=\frac{2\pi^2}{45}g_*T_0^3\simeq 2891.2/$cm$^3$ is the present value of the entropy density, and $\rho_{cr}=1.054\times 10^{-5}h^2$GeV/cm$^3$ the critical density. Equation (\ref{DM1}) must reproduce the observed DM abundance \cite{Planck}
 \begin{equation}\label{DM2}
   \Omega_{DM}h^2\Big|_{obs}=0.1188\pm 0.0010\,,
 \end{equation}
and at the same time, explain the IceCube data.

In the case of the 4-dimensional operator (\ref{4dop}), the dominant contributions to DM production are $a)$ the {\it inverse decay} processes $\nu_\alpha+H^0\to \chi$ and $l_\alpha + H^+\to \chi$, that occurs when $m_\chi> m_H+m_{\nu, l}$ proportional to factor $|y_{\alpha \chi}|^2$, and $b)$ the {\it Yukawa production} processes, such as $t+{\bar t}\to {\bar \nu}_\alpha+\chi$ is proportional to  $|y_{\alpha \chi} y_t|^2$, where $t$ represents the quark top. For the 4-dimensional operator one gets
 \begin{equation}\label{dYtot}
 \frac{dY_\chi}{dT}=\frac{dY_\chi}{dT}\Big|_{inv.dec.}+\frac{dY_\chi}{dT}\Big|_{Yuk.prod.}\,,
 \end{equation}
whose explicit expressions are
 \begin{eqnarray}
   \frac{dY_\chi}{dT}\Big|_{inv.dec.} &=& -\frac{m_\chi^2 \Gamma_\chi}{\pi^2 Hs}\, K_1\left(\frac{m_\chi}{T}\right)\,, \label{dYinvdec} \\
   \frac{dY_\chi}{dT}\Big|_{Yuk.prod.} &=& -\frac{1}{512\pi^6Hs}\int d{\tilde s}d\Omega \sum_\alpha
   \frac{W_{t{\bar t}\to{\bar \nu}_\alpha \chi}+2W_{t\nu_{\alpha}\to t\chi}}{\sqrt{\tilde s}}K_1\left(\frac{\sqrt{\tilde s}}{T}\right)\,, \label{dYYukprod}
 \end{eqnarray}
with ${\tilde s}$ the centre-of-mass energy, $\Gamma_\chi$ the interaction rate given by
 \begin{equation}\label{ratechi}
   \Gamma_\chi=\sum_\alpha \frac{|y_{\alpha\chi}|^2}{8\pi}\, m_\chi\,, \qquad \alpha=e, \mu, \tau\,,
 \end{equation}
and $K_1(x)$ is the modified Bessell function of the second kind. Since $\frac{dY_\chi}{dT}\Big|_{inv.dec.}$ is dominant\footnote{It can be shown \cite{merle} that for the range of values of $y_t\in [0.5, 1]$ (such values of $y_t$ covers all possible values obtained by running its value with the energy) and $m_\chi\in [10^4, 10^8]$GeV, one gets $\Omega_{DM}h^2|_{Yuk.prod.}\simeq 10^{-2}\Omega_{DM}h^2|_{inv.dec.}$, hence $\Omega_{DM}h^2|_{obs}\simeq \Omega_{DM}h^2|_{inv.dec.}$, i.e. the DM relic abundance is mainly generated by inverse decay processes.} with respect to $\frac{dY_\chi}{dT}\Big|_{Yuk.prod.}$, one finds that the relic abundance induced by inverse decay term is
 \begin{equation}\label{DMinvdec}
   \Omega_{DM}h^2|_{inv.dec.}=0.1188 \left(\frac{106.75}{g_*}\right)^{3/2}\frac{\sum_\alpha |y_{\alpha \chi}|^2}{7.5\times 10^{-25}}\,.
 \end{equation}
From (\ref{DMinvdec}) immediately follows that to have the correct DM relic abundance (\ref{DM2}) one has to require
 \begin{equation}\label{ychi}
   \sum_{\alpha=e, \mu,\tau}|y_{\alpha\chi}|^2=7.5\times 10^{-25}\,.
 \end{equation}
However, Eq. (\ref{DMinvdec}) is in conflict with the value of $\sum_{\alpha=e, \mu,\tau}|y_{\alpha\chi}|^2$ needed to explain the IceCube data. To see that, first note that the DM lifetime $\tau_\chi=\Gamma_\chi^{-1}$ has to be larger that the age of the Universe, $\tau_\chi> t_U\simeq 4.35\times 10^{17}$sec. Moreover, IceCube spectrum sets a constraints on lower bounds of DM lifetime $\tau_\chi^b\simeq 10^{28}$sec, i.e. $\tau_\chi \gtrsim \tau_\chi^b$, which is (approximatively) model-independent (see \cite{merle}). Inserting (\ref{ychi}) into (\ref{ratechi}) one obtains
 \[
 \Gamma_\chi\simeq 4.5 \times 10^4 \frac{m_\chi}{\text{1PeV}}\text{sec}^{-1} \quad \to \quad
 \tau_\chi \simeq 2.2\times 10^{-5} \frac{\text{1PeV}}{m_\chi}\text{sec}\ll t_U\,.
 \]
However, the observations of   IceCube  require the dark matter decay lifetime $\tau_\chi= \sim 10^{28}$ sec which implies
 \begin{equation}\label{yIceCube}
   \sum_{\alpha}| y_{\alpha\chi}|^2\simeq 10^{-58}\,,
 \end{equation}
and which is $\sim 33$ order of magnitudes smaller than the value of $\sum_{\alpha=e,\mu, \tau}|y_{\alpha\chi}|^2\sim 10^{-25}$ needed to explain the DM relic abundance, see (\ref{ychi}). As a consequence, the IceCube high energy events and the DM relic abundance are not compatible with the DM production if the latter is ascribed to the 4-dimensional operator $\overline{ L_{L_\alpha}}H\chi$.

\section{PeV neutrinos in modified cosmologies}

As discussed in the previous Sections, the 4-dimensional operator fails to explaining both the IceCube data and DM relic abundance. This is also a consequence of the assumption that the early cosmological background evolves according to GR. The characteristics of the Universe expansion, such as the expansion rate and the composition, affects the relic energy density of DM, as well as their velocity distributions before structure formation. According to the standard cosmological model, the computation of the relic density of particles relies on the assumption that the radiation dominated era began before the main production of relics (and that the entropy of matter is conserved). However, any contribution to the energy density (in matter and {\it geometrical} sector)  modifies the Hubble expansion rate, hence the relic density.

In modified cosmologies (MC), the expansion rate of the Universe can be rewritten in the form \cite{fornengo,BD}
 \begin{equation}\label{H=AHGRIce}
   H_{MC}(T)=A(T)H_{GR}(T)\,,
 \end{equation}
where $A(T)$ is the so called (de)amplification factor. To preserve the successful predictions of BBN, one refers to the pre-BBN epoch since it is not directly constrained by cosmological observations. This means $A(T) \neq 1$ at early time, and $A(T)\to 1$ before BBN begins. Typically the (de)amplification factor can be parameterized as
 \begin{equation}\label{A(T)Ice}
   A(T)=\eta\left(\frac{T}{T_*}\right)^\nu\,,
 \end{equation}
where $T_*$ is a reference temperature, and $\{\eta, \nu\}$ free parameters that depend on the cosmological model under consideration\footnote{For example, in \cite{fornengo} the enhancement function $A(T)$ is parameterized as
 \begin{equation}\label{A(T)}
    A(T)=\left\{ \begin{array}{lcr}
    1+\eta\left(\frac{T}{T_f}\right)^\nu \tanh \frac{T-T_{re}}{T_{re}} & \mbox{for} & T>T_{BBN} \\
    1 & \mbox{for} & T\leq T_{BBN} \end{array} \right.
 \end{equation}
where $T_{BBN}\sim 1$MeV. In the regime $T\gg T_{BBN}$, the function (\ref{A(T)}) behaviors as (\ref{A(T)Ice}). $T_f$ is the temperature at which the WIMPs DM freezes-out, $T_f \simeq 10$GeV.}. Investigations  along these lines have been performed in different cosmological scenarios \cite{fornengo,BD,gondolo}, where The parameter $\nu$ labels cosmological models: $\nu=2$ in Randall-Sundrum type II brane cosmology \cite{randal}, $\nu=1$ in kination models \cite{kination}, $\nu=0$ in cosmologies with an overall boost of the Hubble expansion rate \cite{fornengo}, $\nu=-0.8$ in scalar-tensor cosmology \cite{fornengoST,fornengo}, $\nu=2/n-2$ in $f(R)$ cosmology, with $f(R)=R+\alpha R^n$ \cite{lamb}.

In terms of the modified expansion rate (\ref{A(T)Ice}), it then follows that the inverse decay processes (\ref{dYinvdec}) takes the form
\begin{equation}\label{dYMC}
\frac{dY_\chi}{dT}\Big|_{inv.dec.} = -\frac{m_\chi^2 \Gamma_\chi}{\pi^2 H_{MC}s}\, K_1\left(\frac{m_\chi}{T}\right)\,,
\quad x\equiv \frac{T}{T_*}\,,
\end{equation}
where
 \begin{equation}\label{Heta}
  H_{MC}s=\frac{3.32\pi^2}{45}g_*^{3/2}\eta\left(\frac{m_\chi}{T_*}\right)^\nu \frac{1}{x^{5-\nu}}\,.
 \end{equation}
By inserting (\ref{dYMC}) and (\ref{Heta}) into (\ref{DM1}) and using $\displaystyle{\int_0^\infty dx x^{3+\nu}K_1(x)=2^{2+\nu}\Gamma\left(\frac{5+\nu}{2}\right)\Gamma\left(\frac{3+\nu}{2}\right)}$, where $\Gamma(z)$ are the Gamma functions, one obtains
 \begin{eqnarray}
   \Omega_{DM}h^2 &=& \frac{45h^2}{1.66\pi^2 g^{3/2}}\frac{s_0 M_{Pl}}{\rho_{cr}}\frac{\Gamma_\chi}{m_\chi}\frac{2^{2+\nu}}{\eta}\left(\frac{T_*}{m_\chi}\right)^\nu
   \Gamma\left(\frac{5+\nu}{2}\right)\Gamma\left(\frac{3+\nu}{2}\right) \label{DMeta} \\
  &\simeq & 0.1188 \left(\frac{106,7}{g_*}\right)^{3/2}\frac{\sum_\alpha |y_{\alpha\chi}|^2}{7.5\times 10^{-24}}\, \Pi\,,
  \end{eqnarray}
where $\Pi$ accounts for all corrections induced by modified cosmology
 \begin{equation}\label{Phi}
  \Pi\equiv \frac{2^{3+\nu}}{3\pi \eta}\left(\frac{T_*}{m_\chi}\right)^\nu
   \Gamma\left(\frac{5+\nu}{2}\right)\Gamma\left(\frac{3+\nu}{2}\right)\,.
 \end{equation}
The above result implies $\nu>-3$ and we have used $\Gamma\left(\frac{5}{2}\right)\Gamma\left(\frac{3}{2}\right)=\frac{3\pi}{2}$. To explain the DM relic abundance and the IceCube data, we have to require
 \begin{equation}\label{Phivalue}
   \Pi\simeq 7.5\times 10^{34}\,.
 \end{equation}
A comment is in order. The general analysis performed in \cite{fornengo} provides upper bound on $\eta$ for the cosmological models with $\nu=-0.8, 0, 1, 2$, i.e. $\eta\lesssim 10\div 10^6$ for DM masses $m_\chi\sim (10^2\div 10^4)$GeV. However, these bounds were derived to explain the PAMELA experiment on the observed electron/positron excess. Relaxing them, the parameters $\{\eta, \nu, T_*\}$ are arbitrary and may be choose such that the condition (\ref{Phivalue}) is fulfilled. Therefore we may have $-3 < \nu <0$ for $T_*< M_\chi$ or $\nu>0$ for $T_*> m_\chi$.

\section{Examples of modified cosmologies}

As pointed out in the Introduction, cosmological observations have provided evidences of cosmic acceleration of the present Universe. Instead to invoke the existence of DE, modifying hence the matter sector of GR, an alternative possibility is to modify/generalize the geometrical sector of GR. This approach leads to ETG, and one of the consequences of dealing with alternative cosmologies is that the thermal history of particles turns out to be modified as compared with GR, Eq. (\ref{H=AHGRIce}).

We shall assume that the Universe is described by a flat Friedman-Robertson-Walker metric
\begin{equation}\label{FRW}
  ds^2 = dt^2-a^2(t)(dx^2+dy^2+dz^2)\,,
\end{equation}
where $a(t)$ is the scale factor. We refer to a Universe radiation dominated, so that the energy density is given by $\rho=\frac{\pi^4 g_* T^4}{30}$, $g_*=106$, while the pressure is $p=\rho/3$ (the adiabatic index is $w=1/3$). The dot will stand for the derivative with respect to the cosmic time $t$.

\subsection{Scalar Tensor Theories (STTs)}

The total action of a STT of gravity is given by $S=S_{STT}+S_m$ \cite{fornengo}, where
 \begin{equation}\label{SSTT}
   S_{STT}=\frac{1}{16\pi} \int d^4 x\sqrt{-{\tilde g}} \left[\Phi^2 {\tilde R}({\tilde g})+4\omega(\Phi){\tilde g}^{\mu\nu}\partial_\mu \Phi \partial_\nu  \Phi-4{\tilde V}(\Phi)\right]\,,
 \end{equation}
and $S_m=S_m[\Psi, {\tilde g}_{\mu\nu}]$ is the matter action (the matter fields $\Psi_m$ couple to the metric tensor ${\tilde g}_{\mu\nu}$). The action (\ref{SSTT}) encodes the Brans-Dicke theory of gravity  for $\omega(\Phi)=\omega=constant$. In the form (\ref{SSTT}), the STT action is refereed as Jordan frame. By means of the conformal transformation ${\tilde g}_{\mu\nu}=A_C(\phi) g_{\mu\nu}$ ($A_C$ is the conformal factor that depends on $\phi(x)$) and setting $\Phi^2=8\pi M_*/A_C^2$, $V(\phi)=A_C^4(\phi){\tilde V}(\phi)/4\pi$, and $\alpha(\phi)=\frac{d\log A_C(\phi)}{d\phi}$ $(=(\omega(\Phi)+3)^{-1}$, the action (\ref{SSTT}) can be casted in the so-called Einstein Frame (EF)
 \begin{equation}\label{SSTTE}
   S_{STT}=\frac{M_*^2}{2}\int d^4 x \sqrt{-g}\left[R(g)+g^{\mu\nu}\partial_\mu \phi\partial_\nu \phi-\frac{2}{M_*^2}V(\phi)\right]\,,
 \end{equation}
while the action of matter fields assumes the form $S_m=S_M[\Psi_m, A_C^2(\phi) g_{\mu\nu}]$. $M_*$ accounts for the fact that the gravitational {\it constant} may vary with the scalar field, $M_*=M_{Pl}(\phi)=G^{-1}(\phi)$. In the FRW flat Universe (\ref{FRW}), the cosmological field equations read
 \begin{eqnarray}\label{HSTT}
   H^2\equiv \frac{{\dot a}^2}{a^2} &=& \frac{1}{3M_*^2}\left[\rho+\frac{M_*^2}{2}{\dot \phi}^2+V(\phi)\right]\,, \\
   \frac{\ddot a}{a}&=& -\frac{1}{6M_*^2}\left[\rho+3p+2M_*^2 {\dot \phi}^2-2V(\phi)\right]\,, \label{accSTT}\\
   {\ddot \phi} &+& 3H{\dot \phi}+\frac{1}{M_*^2}\left[\frac{\alpha(\phi)}{\sqrt{2}}(\rho-3p)+V_\phi\right]=0\,, \label{phiSTT}
 \end{eqnarray}
where $V_\phi=\partial V/\partial\phi$. The Bianchi identity (conservation of the energy momentum tensor) $d(\rho a^3)+p d(a^3)=(\rho-3p)d\log A_C(\phi)$ implies $Ta=constant$ for $w=1/3$. From Eqs. (\ref{HSTT})-(\ref{phiSTT}), one gets \cite{fornengo}
 \begin{equation}\label{H2AHGR2}
   H^2\equiv H_{MC}^2=\frac{A_C^2(\phi)[1+\alpha(\phi)\phi^{\,\prime}]^2}{1-\phi^{\,\prime\, 2}/6}H_{GR}^2\,.
 \end{equation}
The prime indicates the derivative with respect to $N\equiv \ln a$ ($\phi^{\,\prime}\equiv\frac{d\phi}{dN}= \frac{d\phi}{d\ln a}=-T\phi_T$, where $\phi_T\equiv d\phi/dT$), while for the scalar field equation one gets (setting $\lambda=V(\phi)/\rho$)
 \begin{equation}\label{phiEq}
   \frac{2(1+\lambda)}{3(1-\phi^{\,\prime\, 2}/6)}\,\phi^{\prime\prime}+[(1-w)+2\lambda]\phi^\prime +\sqrt{2}
   \alpha(\phi)(1-3w)+2\lambda \frac{V_\phi}{V}=0\,.
 \end{equation}
The form of the factor $A(T)$ for a STT follows from (\ref{H2AHGR2})
 \begin{equation}\label{ATSTT}
   A(T)\equiv \frac{A_C(\phi)[1+\alpha(\phi)\phi^\prime]}{(1-\phi^{\,\prime\, 2}/6)^{1/2}}\,.
 \end{equation}
Assuming that $A(T)$ is of the form (\ref{A(T)Ice}), Eq. (\ref{ATSTT}) can be rewritten in the form
 \begin{equation}\label{EqD1}
   \left[A_C^2\alpha^2T^2+\frac{\eta^2 T_*^2}{6}\left(\frac{T}{T_*}\right)^{2(\nu+1)}\right]\phi_T^2-2A_C^2 \alpha\phi_T+\left[A_C^2-\eta^2\left(\frac{T}{T_*}\right)^{2\nu}\right]=0\,.
 \end{equation}
To solve Eq. (\ref{EqD1}) one has to specify the form of $A_C(\phi)$. We study some particular cases:
 \begin{itemize}
   \item $\alpha, A_C \ll 1$ - In this regime Eq. (\ref{EqD1}) reduces to the form
    $\displaystyle{\left(\frac{d\phi}{dz}\right)^2=\frac{6}{z^2}}$, where $z\equiv \displaystyle{\frac{T}{T_*}}$,
    whose solution is $\phi(z)=\phi^{(0)}+ \sqrt{6}\log z \simeq \phi^{(1)}-\sqrt{6}N$, with $\phi^{(0)}, \phi^{(1)}$ constants. Noting that $\phi''=0$, Eq. (\ref{phiEq}) assumes the form $-\frac{\sqrt{6}}{3}+\frac{V}{\rho}(1+\log V_\phi)=0$. Writing the energy density in terms of the field $\phi$, $\rho=K_* e^{4\phi/\sqrt{6}}$ with $K_*\equiv \frac{\pi^2 g_* T_*^4}{30}$,
   Eq. (\ref{phiEq}) allows to derive the potential $V$ (the integration constant is set equal to zero) $V(\phi)=\gamma^{-1}K_*e^{\gamma\phi}$, where $\gamma\equiv \frac{\sqrt{6}}{\sqrt{6}+4}$. The potential $V$ is suppressed for temperatures $T<T_*$.
   \item $\alpha \gg 1$ - In this case Eq. (\ref{EqD1}) becomes
    \[
    \alpha T^2\phi_T - 2=0\,,
    \]
    that gives $\int \alpha(\phi) d \phi=-2/T$, which implies $A_C(\phi(T))=A_0 e^{-2T_r/T}=A_1e^{-2a/a_1}$, where $(A_0, A_1)$ and $(T_r, a_r)$ are integration constants. Therefore the conformal factor diminishes for decreasing (increasing) temperature (scale factor). Consistently with our assumption $\alpha\gg 1$, we must require $\frac{d\phi}{da}=-\frac{2T_r}{\alpha}\ll 1$, so that from (\ref{phiEq}) it follows that $V(\phi)\sim V_0$, where $V_0$ is a constant.
 \end{itemize}
In these examples, results are independent on $\nu$ and $\eta$. To obtain the correct DM relic abundance and explain the IceCube results, hence to fulfill the condition (\ref{Phivalue}), the parameters $\{\eta, \nu, T_*\}$ must fine tuned. Setting $\eta\sim {\cal O}(1)$ and $T_*=10^{q}$GeV, and for $T_*\gg m_\chi$, Eq. (\ref{Phivalue}) implies $\nu=\displaystyle{\frac{34}{q-6}}$. For example, if the transition temperature occurs at $\sim 10^{12}$GeV, i.e. $q\sim 12$, then it follows $\nu\sim 5-6$.

\subsection{Brans-Dicke theory}

In this Section we consider Brans-Dicke (BD) theory of gravity. The BD action follows from the most general action \cite{reviewBD,BransDicke}
 \begin{equation}\label{BDGeneral}
   S=\int d^4x \sqrt{-g}\left[\phi R-\frac{\omega(\phi)}{\phi}\nabla_\mu \phi \nabla^\mu \phi-V(\phi)\right]\,,
 \end{equation}
when $\omega\to constant$ and $V\to 0$. Here $\omega(\phi)$ is an arbitrary function (the coupling parameter).
In Ref. \cite{576BD,reviewBD,105BD} it was shown that during the radiation dominated era, the solutions of the field equations are of the form
\begin{eqnarray}\label{aBD}
  a(\tau) &=& a_0 c (\tau+\tau_+)^{\frac{1}{2}+\alpha}(\tau+\tau_-)^{\frac{1}{2}-\alpha}\\
  \phi(\tau) &=& \phi_0 (\tau+\tau_+)^{-\alpha}(\tau+\tau_-)^{\alpha}\,, \label{phiBD}
\end{eqnarray}
for $\omega>-3/2$, and
\begin{eqnarray}\label{aBD2}
 a(\tau) &=& a_0 \sqrt{(\tau+\tau_-)^2+\tau_+^2}\, e^{-\beta}\,, \\
 \phi(\tau) &=& \phi_0 \beta\,, \label{phiBD2}
\end{eqnarray}
for $\omega<-3/2$. Here $\tau$ is the conformal time, related to the cosmic time by the relation $t=\int a(\tau) d \tau$, $\tau_+$, $\tau_-$, $a_0$ and $\phi_0$ are arbitrary integration constants such that $\frac{8\pi \rho_{rad0}}{3a_0^2\phi_0}=1$, and
 \[
 \alpha \equiv \frac{1}{2\sqrt{1+\frac{2\omega}{3}}}\,, \qquad \beta\equiv \frac{1}{\sqrt{\frac{2|\omega|}{3}-1}}\,
 \tan^{-1}\frac{\tau+\tau_-}{\tau_+}\,.
 \]
The interesting aspect of these solutions is that for late time the scale factor becomes $a(\tau)\sim \tau \sim t^{1/2}$, $\phi\to \phi_0$, i.e. the standard cosmological model is recovered. As an example to explain the IceCube data and the DM relic abundance, we shall consider the solution (\ref{aBD}). Writing the expansion rate in the form (\ref{H=AHGRIce}), $H_{MC}(\tau)=A(\tau)H_{GR}(\tau)$ ($H_{GR}=\frac{1}{\tau}$), we get
 \begin{equation}\label{Atau}
 A(\tau)=\left[\left(\frac{1}{2}+\alpha\right)\frac{\tau}{\tau+\tau_+}+
 \left(\frac{1}{2}-\alpha\right)\frac{\tau}{\tau+\tau_-}\right]\,.
 \end{equation}
Notice $A(\tau)\to 1$ as $\tau\gg \tau_\pm$. To make some estimations, we assume hence that in the early time $\tau < \tau_{\pm}$ (for example, we can set $\tau_- \sim \tau_{BBN}$ and $\tau_+ =\tau_*$ the transition time), so that
 \[
 A(\tau)=\eta \frac{\tau(T)}{\tau_+}\,, \qquad \eta\equiv \left(\frac{1}{2}+\alpha\right)-
 \left(\alpha-\frac{1}{2}\right)\frac{\tau_+}{\tau_-}\,.
 \]
To apply the above result to (\ref{Heta}) we should determine the relation between the conformal time $\tau$ and the temperature $T$. This task cannot be solved analytically. However, we note that whatever is the relation $\tau=\tau(T)$, since $\Pi\sim \eta^{-1}$, to fulfill the condition (\ref{Phivalue}) we can also look at values of parameters for which $\eta\ll 1$. The latter condition implies $\tau_+ \simeq \frac{2\alpha+1}{2\alpha-1}\tau_-$, which requires $\omega>0$.
Of course, the solutions here analyzed are just a subclass of solutions. More general solutions and a richer phenomenology follow, for example, for the general cases in which $\omega(\phi)\neq 0$ and the potential $V(\phi)\neq 0$.


\subsection{$f({\cal T})$ cosmology}

Another interesting model able to explain the accelerated phase of the Universe is provided by the theory of gravity based on the Weitzenb\"{o}ck connection (instead of the usual Levi-Civita connection), and the gravitational field is described by the torsion (instead of the curvature tensor). The torsion tensor is construct in terms of the first derivatives of tetrad fields (no second derivatives appear). This model is referred as the Teleparallel Equivalent of
General Relativity (TEGR), that is  equivalent to General Relativity at the level of field equations \cite{Ferraro2}.
These models represent an alternative to inflationary models, as well as to effective DE models, in which the Universe acceleration is driven by the torsion terms
\cite{Ferraro2,Linder:2010py} (for a detailed review,  see \cite{Pereira.book,Cai:2015emx}). It has been recently discussed in \cite{Cai18}, in the framework of possible future measurement in advancing gravitational wave astronomy, possible tests able to distinguish among modified $f(T)$ gravity.
%

\begin{figure}[btp]
  \centering
  \includegraphics[width=8.0cm]{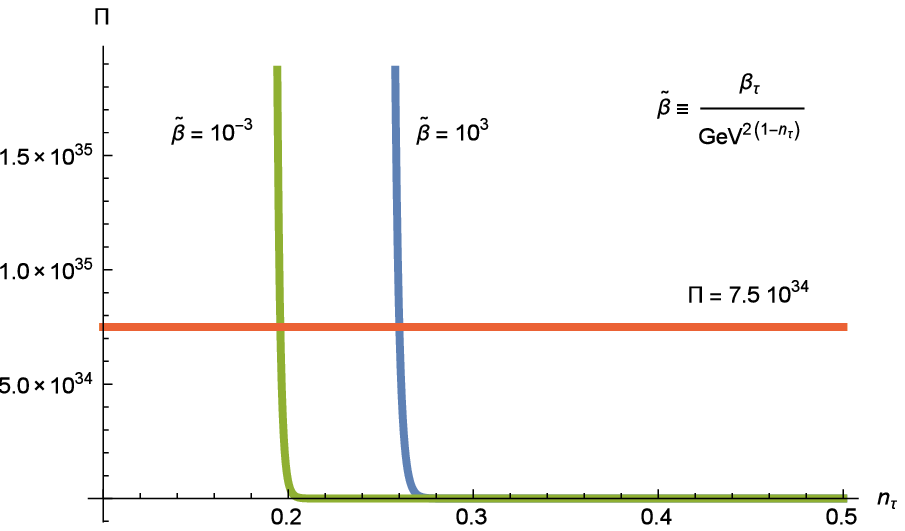}\\
  \caption{$\Pi$ vs $n$. For fixed values of $\eta=1$ and $\beta_{\cal T}$, the value of $\Pi$ needed to explain DM relic abundance and IceCube data follow for $n_{\cal T}\sim 0.25$.}\label{PiTorsion}
   \includegraphics[width=8.0cm]{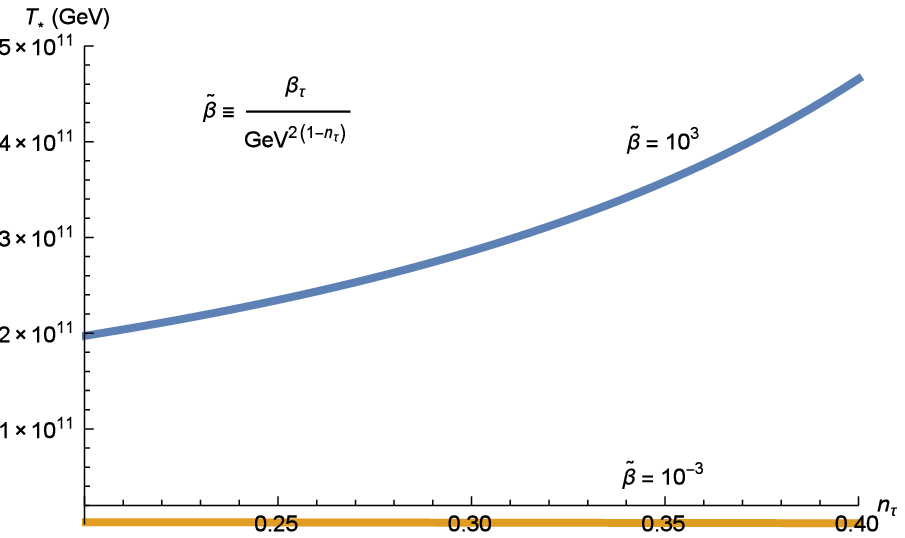}\\
  \caption{$T_*$ vs $n$. For fixed values of $\beta$, the transition temperature occurs at $10^{11}-10^9$GeV, i.e. $T_*\gg m_\chi$.}\label{TTorsion}
  \includegraphics[width=8.0cm]{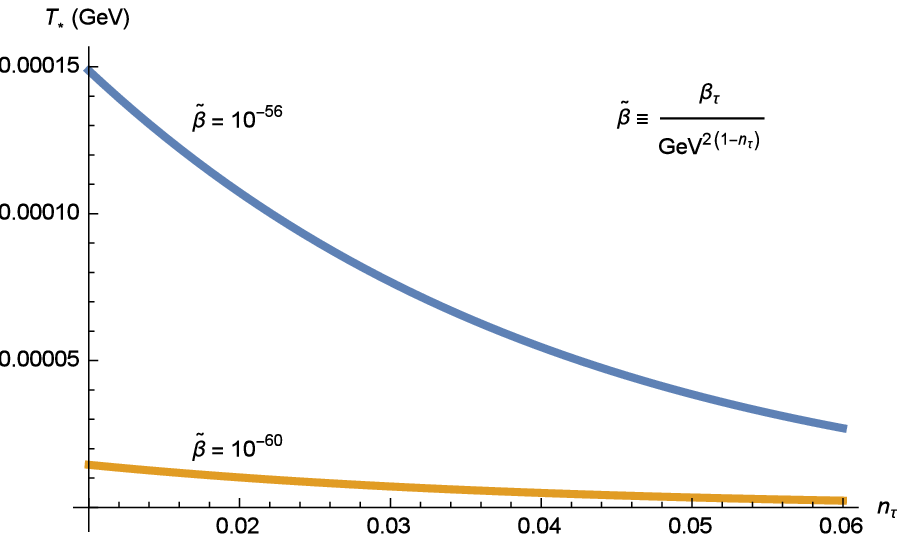}\\
  \caption{$T_*$ vs $n_{\cal T}$. The range of $n_{\cal T}\simeq 0.05$ is taken from the best fit of $CC+H_0+SNeIa+BAO$ data. Here ${\tilde \beta}\sim (H_0/\mbox{GeV})^{2(1-n_{\cal T})}\Omega_{0m}$. The transition temperature occurs at BBN temperature $T_*\sim 0.1$ MeV.}\label{TTorsionCDM}
\end{figure}

In teleparallel gravity, one adopts the curvatureless Weitzenb\"{o}ck connection (that encompasses all the information about the gravitational field)
\begin{equation}\label{torsion}
{\cal T}^\lambda_{\mu\nu}=\hat{\Gamma}^\lambda_{\nu\mu}-\hat{\Gamma}^\lambda_{\mu\nu}
=e^\lambda_i(\partial_\mu e^i_\nu - \partial_\nu e^i_\mu)\,,
\end{equation}
where $e^i_\mu(x)$ are the vierbein fields defined as $g_{\mu\nu}(x)=\eta_{ij} e^i_\mu(x)e^j_\nu(x)$. The action is given by $S^0_I = \frac{1}{16\pi G}\int d^4x e {\cal T}$, where ${\cal T}={S_\rho}^{\mu\nu}{T^\rho}_{\mu\nu}$ is the torsion scalar, $e=det(e^i_\mu)=\sqrt{-g}$, and
\begin{equation}\label{s}
    {S_\rho}^{\mu\nu} = \frac{1}{2}\left[\frac{1}{4}({{\cal T}^{\mu\nu}}_\rho-{{\cal T}^{\nu\mu}}_\rho-{{\cal T}_\rho}^{\mu\nu})
     + \delta^\mu_\rho
{{\cal T}^{\theta\nu}}_\theta-\delta^\nu_\rho {{\cal T}^{\theta\mu}}_\theta\right]\,.
\end{equation}
We shall consider the simplest generalization of the action
$S^0_I$ to construct gravitational modifications based on torsion, i.e.
\begin{equation}\label{action}
    S_I = \frac{1}{16\pi G}\int{d^4xe\left[{\cal T}+f({\cal T})\right]},
\end{equation}
where $f({\cal T})$ is a generic function of the torsion.
For homogeneous and isotropic geometry (\ref{FRW}), the vierbein fields assume the form $e_{\mu}^A={\rm diag}(1,a,a,a)$. By using Eqs. (\ref{torsion}) and (\ref{s}) one infers a relation between the torsion and the expansion rate of the Universe ${\cal T}=-6H^2$. The cosmological field equations read \cite{Cai:2015emx}
\begin{equation}\label{friedmann}
    12H^2[1+f_{\cal T}]+[{\cal T}+f]=16\pi G\rho,
    \end{equation}
    \begin{equation}
     48H^2f_{{\cal T}{\cal T}}\dot{H}-(1+f_{\cal T})[12H^2+4\dot{H}]-({\cal T}-f)=16\pi G p\,,\nonumber  
\end{equation}
where $f_{\cal T}=df/d{\cal T}$. The equations close by taking into account the equation of continuity
$\dot{\rho}+3H(\rho+p)=0$. We consider the power-law $f(T)$ model
\cite{Nesseris:2013jea,Ferraro3}
 \begin{equation}\label{eq:ftmyrzabis}
f({\cal T}) = \beta_{\cal T} |{\cal T}|^{n_{\cal T}},
\end{equation}
By rewriting (\ref{friedmann}) in the form $H^2+H_{\cal T}^2=\displaystyle{\frac{8\pi}{3M_{Pl}^2}\rho}$, where $\displaystyle{H_{\cal T}^2\equiv \frac{f}{6}-\frac{Tf_{\cal T}}{3}}=6^{n-1}\beta_{\cal T}(2n_{\cal T}+1)H^{2n}$, and assuming $H_{\cal T}\gg H$, one gets $H_{\cal T}\equiv H_{MC}=A(T)H_{GR}$, where $A(T)$ is of the form (\ref{A(T)}) with
 \begin{equation}\label{ATTorsion}
   \eta=1\,, \quad \nu=\frac{2}{n_{\cal T}}-2\,, \quad T_*\equiv \left(\frac{24\pi^3 g_*}{45}\right)^{\frac{1}{4}}
   (2n_{\cal T}+1)^{\frac{1}{4(1-n_{\cal T})}}\left(\frac{\beta_{\cal T}}{\mbox{GeV}^{2(1-n_{\cal T})}}\right)^{\frac{1}{4(1-n_{\cal T})}}\left(\frac{M_{Pl}}{\mbox{GeV}}\right)^\frac{1}{2}\mbox{GeV}\,.
 \end{equation}
It is straightforward to show that for the above solution and $a(t)=a_0 t^\delta$, i.e. $H=\frac{\delta}{t}$, it follows  $T(t)a(t)=constant$.
The transition temperature $T_*$ given in (\ref{ATTorsion}) (following from $H_{\cal T}(T_*)\simeq H_{GR}(T_*)$) has to be used into Eqs. (\ref{Phi}) and (\ref{Phivalue}). In Fig. \ref{PiTorsion} we plot $\Pi$ vs $n$. The value of $\beta_{\cal T}$ is obtained by fixing the transition temperature at $T_*\sim 10^{11}-10^9$GeV (see Fig. \ref{TTorsion}), that is $T_*\gg m_\chi$. The parameter $\delta$ enters into the expression of $p$ (we shall not present explicitly being not relevant for our analysis).

For completeness, we also discuss the possibility to use the best fit of the parameters $\{\beta_{\cal T}, n_{\cal T}\}$ for explain the observed accelerated phase of the Universe. This is obtained from the $CC+H_0+SNeIa+BAO$ observational data \cite{TJCAP16} and give $\beta_{\cal T}=(6H_0^2)^{1-n_{\cal T}}\frac{\Omega_{m0}}{2n_{\cal T}-1}$ and $n_{\cal T}=0.05536$, where ${\displaystyle \Omega_{m0}=\frac{8\pi G \rho_{m}}{3H_0^2}}$ is the matter
density parameter at present, and $H_0 = 73.02\pm 1.79\mbox{km/(sec Mpc)} \sim  2.1 \times 10^{-42}\mbox{GeV}$
is the current Hubble parameter. In Fig. \ref{TTorsionCDM} are reported results by fixing $n_{\cal T}\simeq 0.055$. We get a transition temperature at BBN era, $T_*\sim 0.1$MeV, provided ${\tilde \beta}\equiv \frac{\beta_{\cal T}}{\mbox{GeV}^{2(1-n_{\cal T})}}\sim 10^{-56}-10^{-60}$. These values do not match the best fit for $\beta_{\cal T}\sim (H_0/\mbox{GeV})^{2(1-n_{\cal T})}\Omega_{0m}$ by $\sim 20$ order of magnitudes.

\section{Conclusions}

The IceCube collaboration has reported several neutrino events with energies varying from TeV to PeV.
A possible explanation for these events is ascribed to DM particle physics. A viable mechanism for the DM production in the early universe is the freeze-in mechanism. In \cite{merle} it was shown that the minimal  dimension four interaction $y L\cdot H \chi$ fails to explain both DM decay rate required for IceCube and  the correct DM abundance. The lowest dimensional operator which can explain the IceCube decay rate and relic abundance is 6-dimensional operator $\frac{\lambda_{\alpha\beta}\lambda_\gamma^\prime}{M_S^2}\left[\overline{\left(L_{L_{\alpha}}\right)^{\cal C}}i\sigma^2
L_{L\beta}\right](\overline{l_{R\gamma}}\chi)$   \cite{[35]}.
 However, all results are obtained assuming that the cosmological background evolves according to GR fields equations.

In this paper, to reconcile the current bound on DM relic abundance with IceCube data in terms of the 4-dimensional operator, we have followed a different perspective that relates the existence of DM hypothesis with modified theories of gravity. Motivated by cosmological observations by Type Ia Supernovae, CMB radiation, and the large scale structure, according to which the present Universe is in an accelerating phase, new theories beyond GR   have been proposed.
We have shown modified gravity models can explain  the IceCube outputs and at the same time the DM relic abundance observed today in a minimal particle physics model. This because the cosmological field equations based on modified gravity models change the thermal history of particles, so that the expansion rate of the Universe can be written in the form $H(T)=A(T)H_{GR}(T)$, encoding in $A(T)$ the parameters characterizing the model of gravity. Using the particular form of the factor $A(T)$ derived from different cosmological models (we have considered STTs, BD gravity and models related to torsion $f({\cal T}$)), we have solved the Boltzmann equation to get the abundance of DM particles. The latter turns out to be modified by a quantity that does only depend on parameters of the modified cosmological models, and allows to explain, consistently, both the IceCube data and the correct DM abundance $\Omega_DM h^2\sim 0.11$.

Finally, we notice that results derived in this paper allow to exclude some models of modified gravity. For example, considering the $f(R)$ gravity, with $f(R)=R+aR^n$, we have found that  IceCube data and DM relic abundance can be explained provided $n < 1$, However, such a value is not favored by recent Planck release, which require $n>1$ (and in fact the Starobisnky model $n=2$ is one of the favorite candidate for Inflation) \cite{Mohantyrev}. Results here discussed do not allow to distinguish between $f(T)$ and BD  theories of gravity.

\acknowledgments
SM thanks INFN for support.

\end{document}